\documentclass[prl,aps,twocolumn]{revtex4-1}
\usepackage{amsmath}
\usepackage{amsfonts}
\usepackage{graphicx}
\usepackage{color}
\usepackage{xcolor}

\begin{document}
\title {Suppression of the overlap between Majorana fermions by orbital magnetic effects in semiconducting-superconducting nanowires}
\author{Olesia Dmytruk and Jelena Klinovaja}
\affiliation{Department of Physics, University of Basel, Klingelbergstrasse 82, CH-4056 Basel, Switzerland}
\date{\today}

\begin{abstract}
We study both analytically and numerically the role of orbital effects caused by a magnetic field applied along the axis of a semiconducting Rashba nanowire in the topological regime hosting Majorana fermions. We demonstrate that the orbital effects can be effectively taken into account in a one-dimensional model by shifting the chemical potential, and, thus modifying the topological criterion. 
We focus on the energy splitting between two Majorana fermions in a finite nanowire and find a striking interplay between orbital and Zeeman effects on this
splitting.
In the limit of strong spin-orbit interaction, 
we find regimes where the amplitude of the oscillating splitting stays constant or even decays with increasing magnetic field, in stark contrast to the commonly studied case where orbital effects of the magnetic field are neglected. The period of these oscillations is found to be almost constant in many parameter regimes.
\end{abstract}

\maketitle

{\it Introduction.} Majorana fermions (MFs) in condensed matter systems have been at the center of attention over many years.
They have been  predicted to emerge in such systems as
 semiconducting nanowires~\cite{alicea2010majorana,lutchyn2010majorana,oreg2010helical,gangadharaiah2011majorana,
 sticlet2012spin,rainis2013towards,maier2014majorana,sarma2012splitting,
 dominguez2012dynamical,klinovaja2012transition,prada2012transport,leijnse2012introduction,nakosai2013majorana,
 reeg2014zero,bjornson2013vortex,weithofer2014electron,dmytruk2015cavity}, 
 $p$-wave superconductors~\cite{brouwer2011probability,degottardi2013majorana,thakurathi2015majorana,
 scharf2015probing}, 
 graphene-like systems~\cite{klinovaja2012electric,klinovaja2012helical,klinovaja2013giant,Dutreix2014,
 klinovaja2013spintronics,black2012edge,
 san2015majorana,zhou2016ising,kaladzhyan2017formation},
 and chains of magnetic atoms~\cite{nadj2013proposal,klinovaja2013topological,braunecker2013interplay,vazifeh2013self,
pientka2013topological,poyhonen2014majorana}, 
with some of these proposals implemented experimentally~\cite{mourik2012signatures,das2012zero,deng2012anomalous,churchill2013superconductor,albrecht2016exponential,nadj2014observation,
 ruby2015end,pawlak2016probing, rokhinson2012fractional,saldana2017electric}. 
In this work, we focus on Rashba nanowire setups
\cite{lutchyn2010majorana,oreg2010helical}, which have been widely implemented experimentally~\cite{mourik2012signatures,das2012zero,deng2012anomalous,churchill2013superconductor,albrecht2016exponential}. 
The experimental evidence  of  MFs in semiconducting nanowires is based on the observation of emerging zero-bias peaks in the differential conductance as a function of magnetic field applied along the nanowire axis~\cite{mourik2012signatures,das2012zero,deng2012anomalous,churchill2013superconductor}. 

However, at large magnetic fields MFs initially localized at two opposite nanowire ends overlap, resulting in finite-energy fermionic states~\cite{prada2012transport,rainis2013towards,sarma2012splitting,maier2014majorana}. 
 So far, theoretical works have predicted that the energy of these fermionic states should grow exponentially 
with increasing  magnetic field, up to the point where these bound states have crossed the gap and merge with the bulk states~\cite{prada2012transport,sarma2012splitting,rainis2013towards,maier2014majorana}. 
  In contrast to that, transport measurements performed on such nanowires reported the observation of constant or decreasing energy splitting of the MFs as a function of  magnetic field~\cite{churchill2013superconductor,albrecht2016exponential}, which was often used as an argument  against MF interpretation of such data~\cite{liu2017phenomenology}. 
  
  In this work, we resolve this paradox between theory and experiment by taking into account orbital magnetic effects neglected so far and study their effect on the MF energy splitting. In reality, nanowires have a finite diameter, indicating that  orbital effects of the magnetic field may be important. We will show that properly accounting for such orbital effects may explain constant or decreasing amplitude of the MF splitting oscillations in the topological phase. Numerical studies of the topological phase diagram taking into account orbital effects are reported for cylindrical and hexagonal nanowires \cite{lim2013emergence,nijholt2016orbital}. However, so far, not much attention has been paid to the importance of orbital effects for characterization of the energy splitting between MFs.
  
\begin{figure}[t] 
\includegraphics[width=0.9\linewidth]{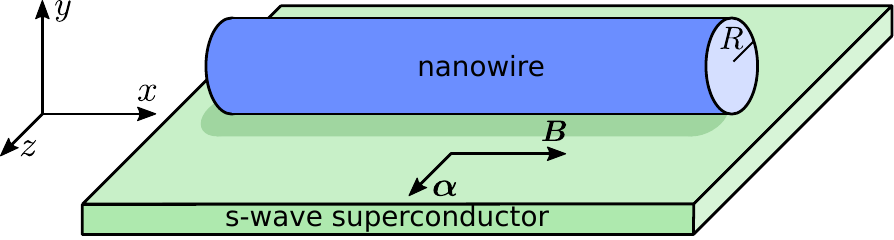}
\caption{The setup consists of the semiconductor Rashba nanowire of radius $R$ brought into proximity to an $s$-wave bulk superconductor. The nanowire is aligned along the $\hat{x}$ axis and the SOI vector $\boldsymbol\alpha$ is along the $\hat{z}$ axis. An external magnetic field $\boldsymbol B$ applied along the nanowire, {\it i.e} along the $\hat{x}$ axis, drives the setup into the topological phase hosting MFs localized at the nanowire ends.} 
\label{fig:scheme}
\end{figure}

In this paper, we propose a one-dimensional (1D) model that takes into account the orbital effects caused by the magnetic field and study how they modify the topological phase. Our system consists of a semiconducting nanowire with Rashba spin-orbit interaction (SOI) in proximity to an $s$-wave bulk superconductor (see Fig.~\ref{fig:scheme}). By applying a magnetic field, such a system can be brought into the topological phase with MFs emerging at the ends of the nanowire. Typically in experiments, the magnetic field is applied along the nanowire axis ($\hat{x}$) in order not to destroy the bulk superconductivity, while the Rashba SOI is orthogonal to the magnetic field $B$ (chosen here along $\hat{z}$). In most theoretical works, the magnetic field is assumed to enter only as a Zeeman term, while the orbital contribution is dismissed due to the small nanowire diameter. As a result, MF oscillations have been found only in the weak SOI regime, where the Fermi wavevector depends on $B$, and thus the amplitude of the splitting always grows  as the localization length grows with increasing $B$-field \cite{prada2012transport,sarma2012splitting,rainis2013towards,maier2014majorana}. In the strong SOI regime \cite{fasth2007direct,nadj2012spectroscopy,van2015spin,
kammhuber2017conductance}, the Fermi wavevector is independent of $B$ unless orbital effects, shifting the chemical potential, are taken into account. The dominant MF localization length,  determined by the proximity  gap at the exterior branches of the wire spectrum, stays constant in this regime, so one expects a constant amplitude of the MF overlap. We demonstrate that, quite remarkably, these orbital effects can qualitatively change the MF energy splitting and thus can account for a better agreement between theory and recent experiments~\cite{albrecht2016exponential}.

{\it Orbital effects in lowest subbands.} A three-dimensional nanowire is described by the Hamiltonian $H_{3D}(x,y,z)$ in which the dynamics along the nanowire ($\hat{x}$-axis) and in the cross-section of the nanowire ($\hat{y}\hat{z}$-plane) are independent,  $H_{3D}(x,y,z) = \bar{H}(x) + H_{2D}(y,z)$. As a result, the wavefunction takes the form $\Psi (x,y,z) = \bar \Psi(x) \Psi_{2D}(y,z)$ and the problem can be solved in two steps. Thus, we first focus on  finding the eigenvalues of $H_{2D}(y,z)$ in the presence of orbital effects. Afterwards, we deal with the effective one-dimensional model, in which the chemical potential $\mu$ shifts as a function of the magnetic field. Below, we show that at small magnetic fields the dependence is quadratic, so $\mu = \mu_0 - \overline{\beta}\left(\Phi/\Phi_0\right)^2$, where $\Phi_0 = hc/e$, $\mu_0$ is the initial chemical potential, $\Phi = B S$ is  the magnetic flux through the nanowire cross-section of area $S$.


The simplest model to consider analytically is a cylindrical hollow nanowire of radius $R$
~\cite{lim2013emergence,winkler2017orbital}. The kinetic term in the transverse direction is written in polar coordinates (in $\hat{y}\hat{z}$ plane, see Fig.~\ref{fig:scheme}) as
\begin{align}
H_{kin}^{cyl} = \int d\phi\ \psi^\dag (\phi) \dfrac{\hbar^2}{2m^*}\left(\dfrac{-i\partial_\phi}{R} - \dfrac{e B R }{2c\hbar}\right)^2 \psi(\phi),
\end{align}
where $m^*$ is the effective mass and the vector potential $A = BR/2$ is chosen in the Coulomb gauge. 
The energy spectrum  is given by $E_l^{cyl} =  \hbar^2 \left(l - \Phi/\Phi_0\right)^2/2m^* R^2$,
where $\Phi = \pi R^2 B$ is the magnetic flux through the cylinder cross-section 
and the quantum number $l$ corresponding to the angular momentum is an integer. In what follows, we work with the lowest non-degenerate subband ($l = 0$), so the orbital effects indeed could be taken into account by shifting the chemical potential up 
proportionally to $B^2$, where the corresponding coefficient  is defined as
$\bar \beta^{cyl} = \hbar^2/2m^* R^2$ (see above). 
\begin{figure}[t!] 
\includegraphics[width=0.9\linewidth]{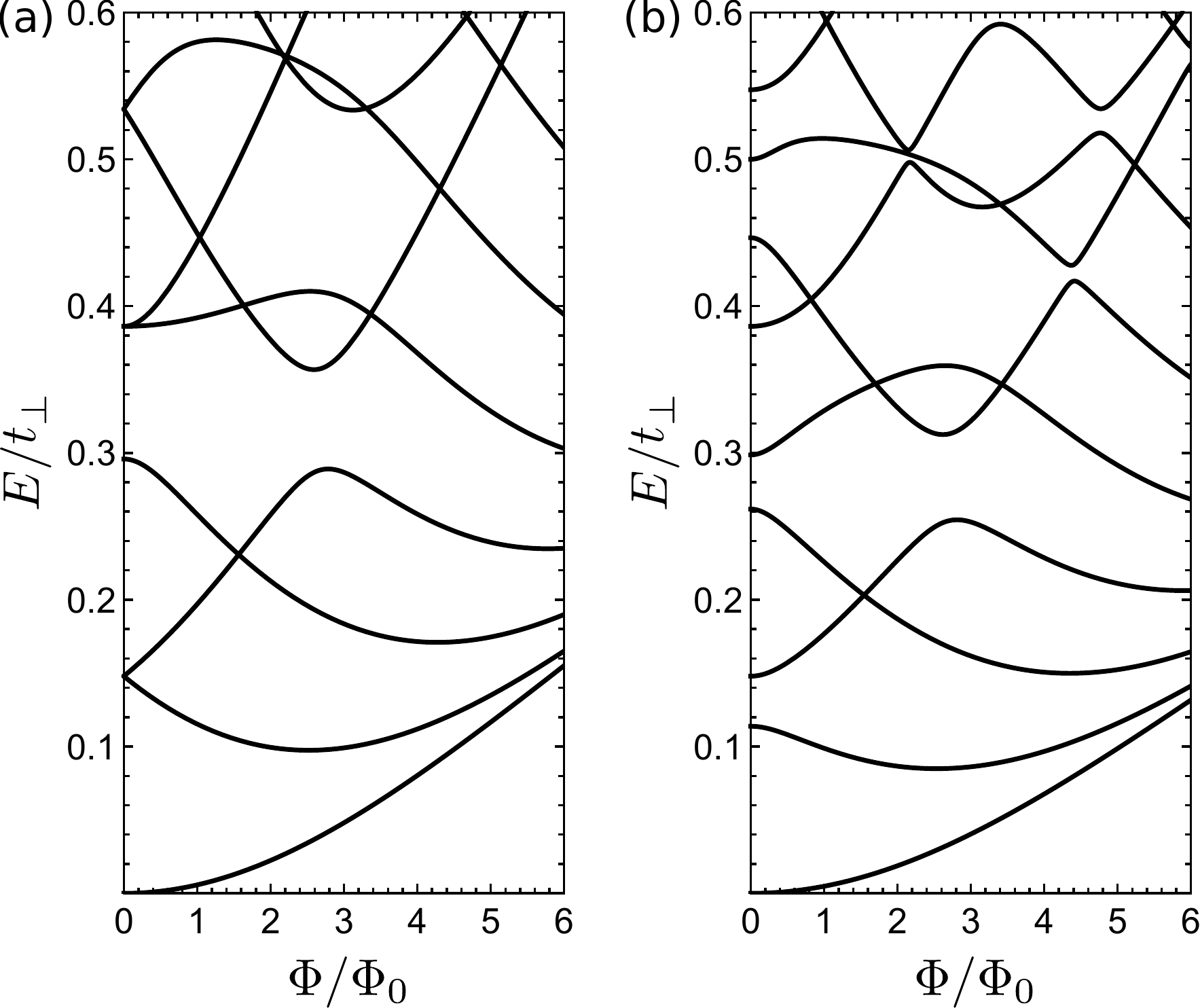}
\caption{The energy spectrum of the nanowire with the cross-section $N_y a \times N_z a$  as a function of the magnetic flux $\Phi/\Phi_0$. 
(a) In a nanowire with a square cross-section  ($N_y \times N_z = 12 \times 12$), the lowest energy level is non-degenerate, while the majority of higher subbands is two-fold degenerate at $B = 0$ due to the additional mirror symmetry. (b) In contrast to that, in a nanowire with a rectangular cross-section ($N_y \times N_z = 14 \times 12$) the symmetry is broken and, as a result, lowest subbands are non-degenerate. In both cases, the bottom of the lowest subband moves up as a quadratic function of $B$.
One flux quantum through the cross-section corresponds to magnetic fields of  strength (a) $B=0.41~T$ or (b) $B=0.35~T$ for $a=8.33~\text{nm}$. The splitting between subbands is determined by $t_\perp = \hbar^2/2m^*a^2= 37$~meV, where $m^*=0.015 m_e$.} 
\label{fig:2Dspectrum}
\end{figure}

Next, we study numerically a more realistic situation in which a nanowire has a rectangular cross section $N_y a \times N_z a$, where $a$ is the effective lattice constant.
In 
the Landau gauge $\vec{A} = B y \hat{z}$, the tight-binding Hamiltonian reads as 
\begin{align}
H_{2D} = 
&- t_\perp \sum_{j = 1}^{N_y} \sum_{k = 1}^{N_z + 1}c^\dag_{j+1,k}c_{j,k} \label{eq:2dHamiltonian} \\ 
&\hspace{35pt}
- t_\perp \sum_{j = 1}^{N_y + 1} \sum_{k = 1}^{N_z} e^{-i\varphi_j} c^\dag_{j,k+1}c_{j,k} +{\rm H.c.},  \nonumber
\end{align}
where the phase $\varphi_j = 2\pi \Phi j / \left(N_y N_z \Phi_0\right)$ accounts for orbital effects, $\Phi = B N_y N_z a^2 $ 
is the magnetic flux though the nanowire cross-section, and $t_\perp>0$ is the hopping amplitude.  
Here, $c_{j,k}^\dag (c_{j,k})$ is the fermionic creation (annihilation) operator at site $(j,k)$ of the square lattice.

In nanowires with a square cross-section $N_y=N_z$, the lowest subband is non-degenerate, while the majority of higher subbands are multiply degenerate at $B=0$ due to the presence of an additional mirror plane going through the square diagonal and the nanowire axis,
see Fig.~\ref{fig:2Dspectrum}(a). This degeneracy should be expected to occur in all nanowires with high-symmetry cross-sections. However, in presence of disorder or working with  nanowires covered only partially by the superconductor, we assume such symmetries are broken and the degeneracy is lifted. For example, if $N_y$ and  $N_z$ are non-commensurable, the lowest energy subbands are non-degenerate [see Fig.~\ref{fig:2Dspectrum}(b)]. Again, for small magnetic fields the bottom of the lowest subband moves up as $\propto B^2$, which is consistent with results obtained for hollow cylinders~\cite{saldana2017electric}.  We note that also the lowest energy levels of the Fock--Darwin spectrum for an electron in a parabolic 2D confinement subjected to small magnetic fields follows the same dependence on the flux~\cite{geerinckx1990effect,madhav1994electronic,burkard1999coupled,tsitsishvili2004rashba}. 
In what follows, we focus on this case of a single non-degenerate band and take into account orbital effects via an effective shift of the chemical potential. 

{\it Effective 1D Hamiltonian.}  Next, we introduce an effective continuum model for a one-dimensional Rashba nanowire described by the Hamiltonian 
 $H = H_{kin} + H_{so} + H_Z+H_{SC}$, where
\begin{align}
 &H_{kin} = \sum_{\sigma}\int dx\ \psi^\dag_\sigma (x)\left[-\hbar^2\partial^2_x/2m^* - \mu\right]\psi_{\sigma}(x),
\label{eq:kinetic}\\
&H_{so} = -i\alpha\sum_{\sigma,\sigma'}\int dx\ \psi^\dag_\sigma (x)\left(\sigma_z\right)_{\sigma\sigma'}\partial_x\psi_{\sigma'}(x),
\label{eq:soi}\\
&H_Z = V_Z\sum_{\sigma,\sigma'}\int dx\ \psi^\dag_\sigma (x)\left(\sigma_x\right)_{\sigma\sigma'}\psi_{\sigma'}(x),
\label{eq:zeeman}\\
&H_{SC} = \dfrac{\Delta}{2}\sum_{\sigma,\sigma'}\int dx\ \psi_\sigma (x)\left(i\sigma_y\right)_{\sigma\sigma'}\psi_{\sigma'}(x) + H.c.
\label{eq:supercond}
\end{align}
with $\alpha$ being the SOI strength, $\Delta$ the proximity-induced pairing gap,
$V_Z=g\mu_B B/2$  the Zeeman energy, where $g$ is the $g$-factor of the nanowire and $\mu_B$  the Bohr magneton.
Here, 
$\psi^\dag_\sigma (x) \left[\psi_\sigma (x)\right]$ is the creation (annihilation) operator of an electron at position $x$ with spin $\sigma/2 = \pm 1/2$, and $\sigma_{x,y,z}$ are the Pauli matrices acting on the spin of the electron.
%
We assume $V_Z$ and  $\Delta$  to be positive without loss of generality.

\begin{figure}[t] 
\includegraphics[width=0.8\linewidth]{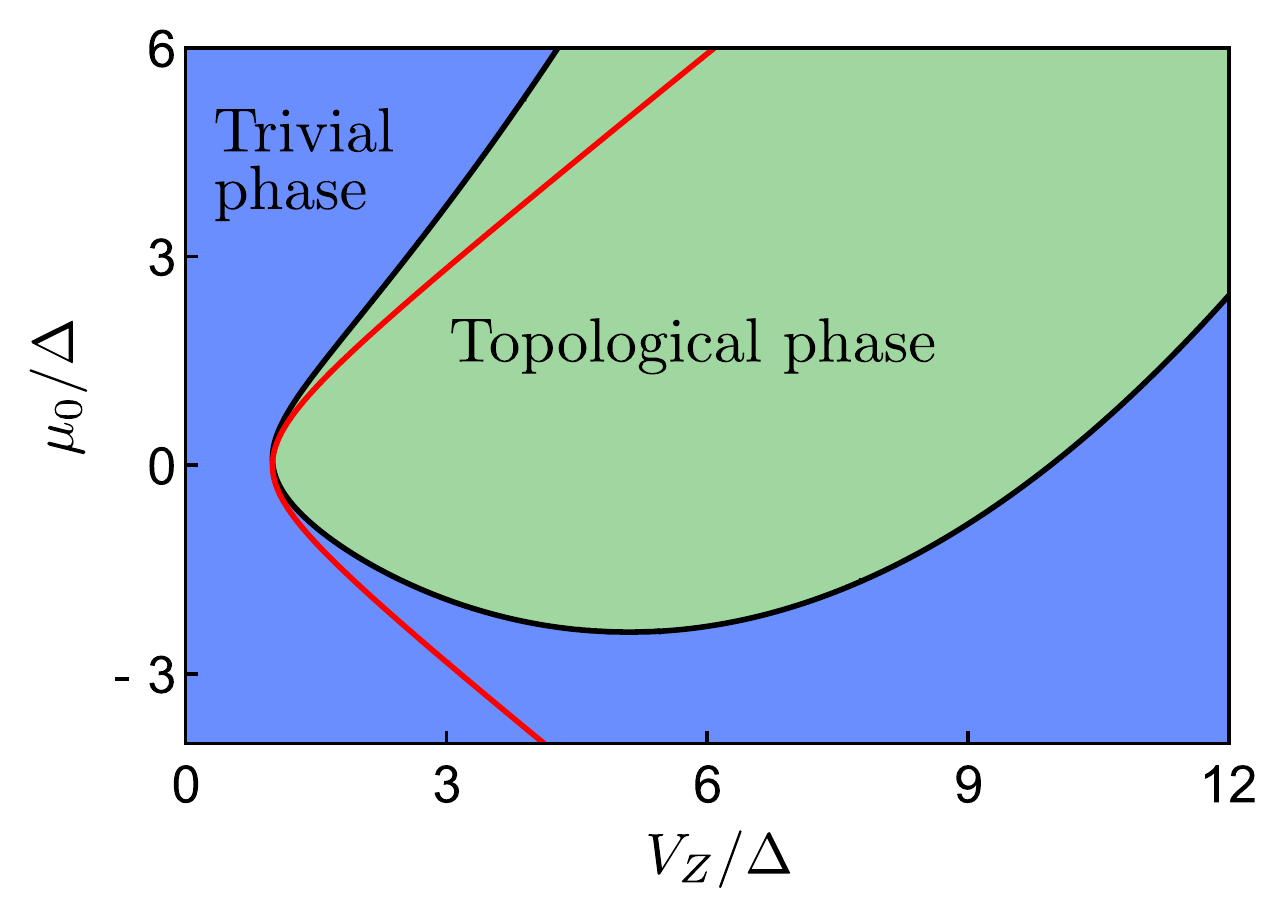}
\caption{Topological phase diagram as a function of applied magnetic field in units of  $V_Z/\Delta$ and of initial chemical potential $\mu_0/\Delta$ for $\beta \Delta = 0.1$. The topological (green area) and trivial (blue area) phases are separated by the phase boundary (black line) corresponding to the closing of the bulk gap. The red line indicates the phase boundary in the absence of orbital effects. Clearly, orbital effects are responsible for shifting the topological phase to higher values of chemical potentials.} 
\label{fig:PhaseDiagram}
\end{figure}

{\it Topological criterion modified by orbital effects.} The topological phase transition is associated with a closing and reopening of the bulk gap. The Rashba nanowire is in the topological phase with MFs appearing at both ends of the nanowire if $V_Z > \sqrt{\mu_0^2+\Delta^2}\equiv V_Z^0$, where the chemical potential $\mu_0$  is calculated from the SOI energy \cite{lutchyn2010majorana,oreg2010helical}.
Orbital magnetic effects taken into account in the effective model as $\mu = \mu_0 - \beta V_Z^2$ [$\beta=\bar \beta \left(2S/g\mu_B\Phi_0\right)^2$] modify the topological criterion.  
In particular, as the magnetic field is increased, the Zeeman energy grows, however, at the same time the bottom of the subband moves up and the chemical potential is decreasing, which makes it more difficult to achieve the topological phase. In particular, if the initial potential is too low, $\mu_0<-1/2\beta$ or $\mu_0<(4\beta^2\Delta^2-1)/4\beta$, the system is always in the trivial phase.
Generally, there are two critical values of magnetic fields $V_{Z,\pm}$ at which  the gap at $k=0$ closes,
\begin{align}
V_{Z,\pm}^2 = (1+ 2 \beta \mu_0  \pm \sqrt{1+4 \beta \mu_0  - 4\beta^2\Delta^2})/2\beta^2,
\end{align}
and, thus, the topological phase transition takes place twice. The topological phase hosting MFs  at each end of the nanowire described by the modified topological criterion $V_{Z,-} < V_Z < V_{Z,+}$. In the limit $\beta\to0$,  we reproduce the standard topological criterion $V_{Z-} = V_{Z}^0$ and $V_{Z+}$ diverges.  If after the first topological phase transition, the magnetic field is increased further, the system could be driven out of the topological phase again (see Fig.~\ref{fig:PhaseDiagram}). In particular, in sufficiently long nanowires, one will observe that the zero-bias MF peak in the conductance disappears without showing any oscillations~\cite{mourik2012signatures}. 
 Moreover, as it is difficult to detect the closing of the bulk gap in the nanowires with a soft superconducting gap via transport measurements, the sudden disappearance of MFs could look puzzling,  if orbital effects are not taken into account. For $(1 - \sqrt{1-\beta^2\Delta^2})/\beta < \mu_0 < (1 + \sqrt{1-\beta^2\Delta^2})/\beta$, $V_{z-}$ is smaller than $V_{Z}^0$ and the topological phase is achieved at smaller magnetic fields. In addition, due to orbital effects, the topological phase shifts towards higher values of chemical potential, which reduces the challenging requirement of tuning the electron density to very low values.
 
{\it MF wavefunctions: semi-infinite nanowire.} After, we have identified the bulk properties, we will focus on MFs in semi-infinite nanowires. To find the MF wavefunctions, we consider the strong SOI regime defined by the condition that the SOI energy is the largest energy scale,  $E_{so}= \hbar^2 k_{so}^2/2 m^* \gg V_Z,\Delta,\mu$, where $k_{so} = m^*\alpha/\hbar^2$. In this regime, 
we linearize the Hamiltonian $H$ near the Fermi points $k^{(i)}_F=0$ ($k_F^{(e)} = \pm  2k_{so}$) corresponding to the interior (exterior) branch of the spectrum with the Fermi velocity $\upsilon_F= \alpha/\hbar $ by expressing the electron operators $\psi_{\sigma}(x)$ in terms of slowly varying left $L_{\sigma}$ and right $R_{\sigma}$ movers~\cite{braunecker2010spin,klinovaja2012composite,klinovaja2012transition}, $\psi_{\uparrow}(x) = R_{\uparrow}(x) + e^{- 2 i k_{so} x} L_{\uparrow}(x)$ and $\psi_{\downarrow}(x) = e^{2 i k_{so} x} R_{\downarrow}(x) +  L_{\downarrow}(x)$.
Next, we construct the basis vector that corresponds to the exterior (interior) branch $\left(\phi^e\right)^T$ = ($L_{\uparrow}$, $R_{\downarrow}$, $L^\dag_{\uparrow}$, $R^\dag_{\downarrow}$)  [$\left(\phi^i\right)^T$ = ($R_{\uparrow}$, $L_{\downarrow}$, $R^\dag_{\uparrow}$, $L^\dag_{\downarrow}$)]. 
The linearized Hamiltonian density, $\tilde{H}^{l} = \frac{1}{2}\int dx\ \left[\phi^l(x)\right]^\dag\mathcal{H}^l\phi^l(x)$,  can be written in terms of Pauli matrices $\eta_{x,y,z}$ acting on the electron-hole subspace as 
\begin{align}
&\mathcal{H}^e = i\hbar\upsilon_F\sigma_z\partial_x + \Delta\sigma_y\eta_y - \mu \eta_z,\nonumber\\
&\mathcal{H}^i = -i\hbar\upsilon_F\sigma_z\partial_x +  V_Z\sigma_x\eta_z + \Delta\sigma_y\eta_y - \mu \eta_z.
\end{align}

Imposing vanishing boundary conditions  at the left end of the nanowire, we find $\Psi_L(x) =
( f, i f^*, f^*, -i f)^T/\sqrt{\mathcal{N}}$ with
\begin{align}
f(x) = \left(-i  e^{i\left(2 k_{so} - \mu/\alpha\right) x -x / \xi^e} + i e^{-x / \xi^i}\right) e^{-i\varphi_L/2},
\label{eq:leftfunction}
\end{align} 
where  $\mathcal{N}$ is the normalization prefactor and $\sin \varphi_L = \mu / V_Z$.
The MF localization lengths are given by $\xi^e = \alpha / \Delta$ and $\xi^i = \alpha / \left(\sqrt{V_Z^2 - \mu^2}-\Delta\right)$.   By analogy, we also find the MF wavefunction localized at the right end of the nanowire and, thus, exponentially decaying to the left, let say, for $x < L$ with $\Psi_R(x = L) = 0$. Not surprisingly, the left and right MF wavefunctions are related as $\Psi_R(x) = \Psi_L^*(L-x)$,
reflecting the mirror  symmetry between the two ends.

\begin{figure}[t!] 
\includegraphics[width=0.85\linewidth]{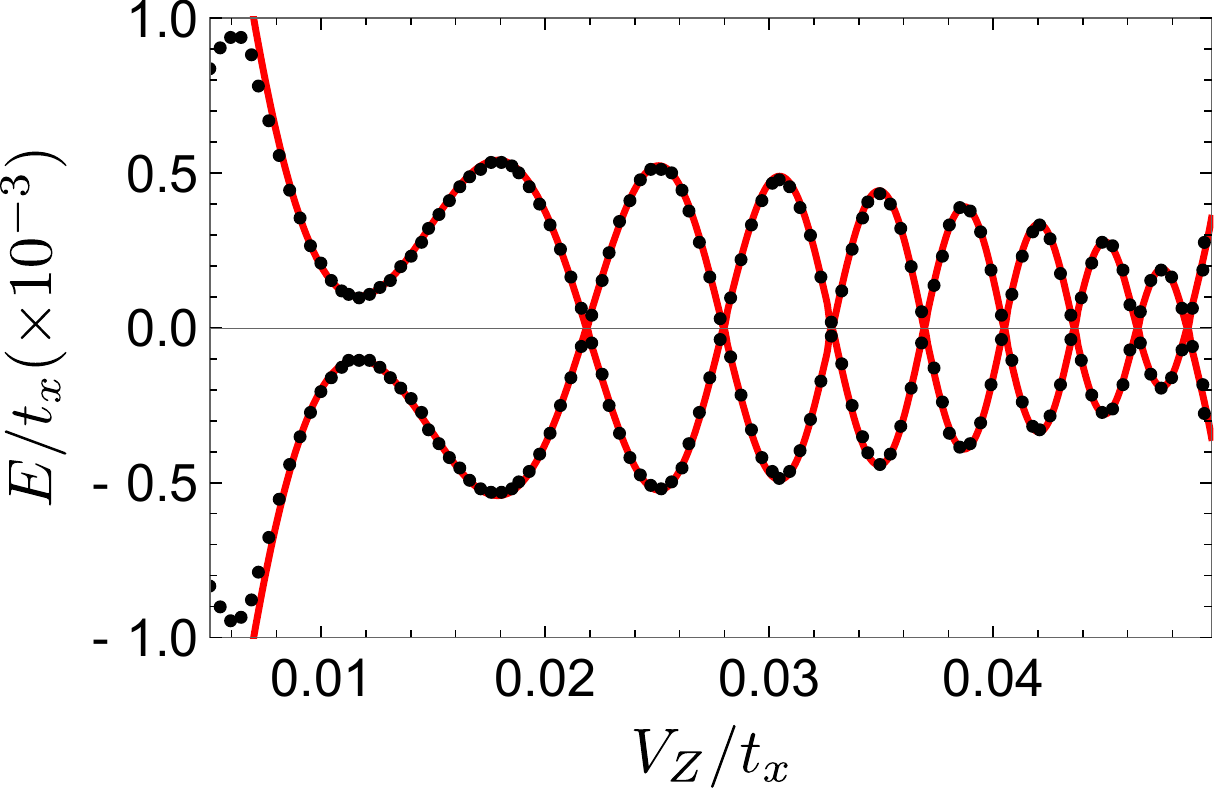}
\caption{ The MF energy splitting as a function of applied magnetic field in units of  $V_Z/t_x$
obtained by numerical diagonalization (red solid line) or using the analytical expression for $\delta\epsilon$ (black dotted line).  The overlap between MFs decays and exhibits oscillations with almost constant period.
The used parameters are  $N = 300$, $\Delta/t_x=0.005$, $\mu = -20 V_Z^2/t_x$ and $\overline{\alpha}/t_x = 0.3$.
}
\label{fig:Energy}
\end{figure}

{\it Tight-binding model.} Next, we would like to focus on the finite-size nanowires and calculate the splitting between two MFs. To achieve this, we first turn to the modeling of the system by using the tight-binding Hamiltonian of a 1D chain composed of $N + 1$ sites~\cite{rainis2013towards,klinovaja2015fermionic} 
\begin{align}
&\overline{H} = \sum_{\sigma,\sigma'}\sum_{ j=1}^{N} 
c^\dag_{j+1,\sigma}\Big[i \overline{\alpha} \sigma^z_{\sigma\sigma'}-t_x \delta_{\sigma\sigma'} \Big]c_{j,\sigma'}- \sum_{j=1}^{N+1} \Delta c^\dag_{j,\uparrow}c^\dag_{j,\downarrow}\nonumber\\
&- \sum_{\sigma,\sigma'}\sum_{j=1}^{N+1} c^\dag_{j,\sigma}\Big[\left(\mu - 2t_x\right) \delta_{\sigma\sigma'}
- V_Z \sigma^x_{\sigma\sigma'}\Big]c_{j,\sigma'}
+{\rm H.c.},
\label{eq:tight-binding_hamiltonian}
\end{align} 
where $c_{j,\sigma}^\dag (c_{j,\sigma})$ is the creation (annihilation) operator acting on electrons with spin $\sigma$ located at  site $j$. Here, $t_x = \hbar^2/\left(2 m^* a_x^2\right)$ is the hopping amplitude along $\hat{x}$, with $a_x$ being the lattice constant, and $\overline{\alpha}$ is the spin-flip hopping amplitude, related to the SOI parameter by $\overline{\alpha} = \alpha/2a_x$. 

\begin{figure}[t!] 
\includegraphics[width=0.8\linewidth]{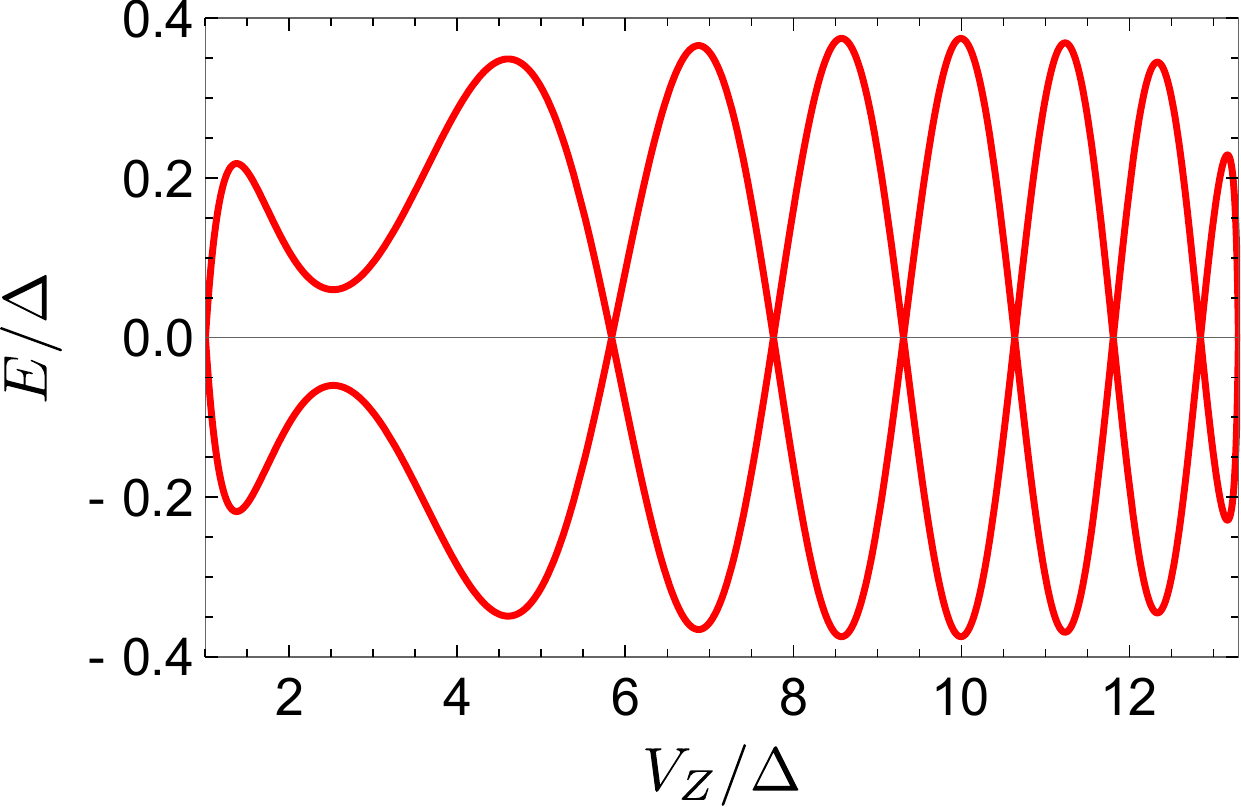}
\caption{The MF energy splitting as a function of applied magnetic field in units of  $V_Z/\Delta$ in finite-size nanowire $k_{so} L = 151$. The amplitude of oscillations stays constant away from the topological phase transition points, close to which it shrinks.
The parameters are chosen as $E_{so}/\Delta = 50$ and  $\mu = - 0.075V_Z^2/\Delta$.} 
\label{fig:Energy2}
\end{figure}

{\it Splitting between MFs.}  Next, we focus on the splitting between MFs. Numerically, we find that the amplitude of MF splitting either stays constant or decays, see Fig. \ref{fig:Energy}.
The left and right MF wavefunctions found independently for a semi-infinite nanowire do not satisfy the Schr\"{o}dinger equation if the nanowire length is finite. Using perturbation theory we find that the degeneracy between the two MF levels is lifted by $\delta\epsilon = \left|\left\langle 0 \left| \gamma_R \overline{H} \gamma_L^\dag \right| 0 \right\rangle\right|$,
where $\gamma_{L,R}$ are MF operators (the details of the derivation are presented in the SM~\cite{SM}). In the regime of strong SOI, $\delta\epsilon$ can be simplified as 
\begin{align}
\delta\epsilon \approx \frac{2\hbar \upsilon_F}{\xi^e + \xi^i}\left|f\left(L\right)\right|
\left|\sin\left(\tilde{\varphi}\right)\right|,
\label{eq:overlap_result}
\end{align}
where $\tilde{\varphi} = \varphi_L/2 - \text{Arg}\left[f(L)\right]$. Away from the topological phase transition point, the exterior gap is the smallest one, so $\xi^e \gg \xi^i$.  As a result, the amplitude of energy splitting $\delta\epsilon$ stays constant and is given by $2 \Delta e^{-L/\xi^e}$. The period of oscillations is given by $\delta V_Z = \pi\alpha / 2 \beta L V_Z$, see Fig.~\ref{fig:Energy2}. 
Close to the topological phase transition point  $\xi^i \gg \xi^e$ and $\delta\epsilon \approx 2\left(\sqrt{V_Z^2 - \mu^2} - \Delta\right)e^{-L / \xi^i}\left|\cos\left(\varphi_L + \theta'\right)\right|$,  where $\theta' \approx e^{L / \xi^i} e^{- L/\xi^e} \sin\left[\left(2 k_{so} - \mu/\alpha\right)L\right]$.  In principle, in this regime we should also get oscillations in $\delta\epsilon$, but this regime is so narrow in the values of the magnetic field due to the the exponential decay, the oscillations are irregular, see Fig. \ref{fig:Energy2}. 

For degenerate bands and also for high values of Zeeman energy, the chemical potential moves linearly as a function of magnetic
field~(see Fig.~\ref{fig:2Dspectrum}). In this case, we observe similar periodic oscillations of the energy splitting between two MFs but the region with shrinking amplitude gets larger due to slower
dependence of $\mu$ on $V_Z$~\cite{SM}. We note that our calculations assumed that the proximity-gap is independent of magnetic fields, which corresponds to the weak coupling regime ~\cite{sau2010robustness,potter2011engineering,kopnin2011proximity,zyuzin2013correlations,
takane2013superconducting,van2016conductance,
reeg2017destructive}.
In the strong coupling  regime \cite{reeg2017finite}, we also took into account effects of external magnetic field on the bulk $s$-wave superconductor  in which the proximity-induced gap $\Delta = \Delta_0\sqrt{1 - \left(V_Z / V_Z^c\right)^2}$ is suppressed at the critical field $V_Z^c$ \cite{SM}. Apart from modifications in the topological criterion, our finding of non-growing oscillations of the MF splitting stays valid also in this case \cite{SM}.

{\it Conclusions.} 
In this work, we take  into account the orbital effects due to the finite-size cross-section of the nanowire by shifting the chemical potential in an  effective 1D model. 
Adding orbital effects leads to modification of the topological phase transition criterion. Moreover, in the strong SOI regime, the amplitude of  the MF energy splitting can stay constant or even decrease as the magnetic field is increased. This result could be relevant for current experimental data~\cite{albrecht2016exponential,churchill2013superconductor}.

\paragraph*{Acknowledgments.}
We thank Daniel Loss and Diego Rainis for motivating discussions. This work was supported by the Swiss National Science Foundation and the NCCR QSIT.

\begin{widetext}

\newpage
\onecolumngrid
\bigskip 

\begin{center}
\large{\bf Supplemental Material to `Suppression of the overlap between Majorana fermions by orbital magnetic effects in semiconducting-superconducting nanowires' \\}
\end{center}
\begin{center}
Olesia Dmytruk and Jelena Klinovaja
\\
{\it Department of Physics, University of Basel, Klingelbergstrasse 82, CH-4056 Basel, Switzerland}
\end{center}

\section{Energy Splitting between two MFs in finite-size nanowire}
In this section we provide details of the calculation of the splitting between two MFs in a finite-size nanowire. We assume that we already found the left ($\Psi_L$) and right ($\Psi_R$) MF wavefunctions~\cite{klinovaja2012c,klinovaja2012t}.  The left MF wavefunction $\Psi_L$ satisfies the Schr\"{o}dinger equation for the semi-infinite nanowire, $\mathcal{H}_0^L \Psi_L = 0$, with corresponding boundary conditions.  The corresponding expressions can be found analytically or numerically by considering the length of the chain $N'a$ to be much larger than the MF localization lengths, see Fig.~\ref{fig:Wavefunction}.  We rewrite $\overline{H}$ in Nambu representation as  matrix $\mathcal H_{ij}$ of  size $4(N+1)\times4(N+1)$ in the basis composed of $(c_{j, \sigma}, c_{j, \sigma}^\dagger)$. By finding eigenvalues and eigenvectors of $\mathcal H_{ij}$, one determines energy levels and corresponding wavefunctions. In the regime of strong SOI, the coefficients $\chi^{L,R}_{j\eta\sigma}$ can be determined from the continuum model considered in the main text.  
Generally, we find good agreement between the two models. The left MF wavefunction for the semi-infinite nanowire [$\Psi_L(n=0)$ is equal to zero in the continuum model] can be written in this basis as 
\begin{align}
\gamma_L = 
\sum_{n= 1}^{\infty}\Big[\left(\chi^{L}_{n11}\right)^*  c_{n,1}+ \left(\chi^{L}_{n1\bar1}\right)^* c_{n,\bar1} + \left(\chi^{L}_{n\bar11}\right)^* c_{n,1}^\dag + \left(\chi^{L}_{n\bar1\bar1}\right)^* c_{n,\bar1}^\dag\Big],
\end{align}
while the corresponding right MF wavefunction [$\Psi_R(n=N+2)$ is equal to zero in the continuum model] reads
\begin{align}
\gamma_R = 
\sum_{n= -\infty}^{N+1}\Big[\left(\chi^{R}_{n11}\right)^*  c_{n,1}+ \left(\chi^{R}_{n1\bar1}\right)^* c_{n,\bar1} + \left(\chi^{R}_{n\bar11}\right)^* c_{n,1}^\dag + \left(\chi^{R}_{n\bar1\bar1}\right)^* c_{n,\bar1}^\dag\Big].
\end{align}

\begin{figure}[b!] 
\includegraphics[width=0.45\linewidth]{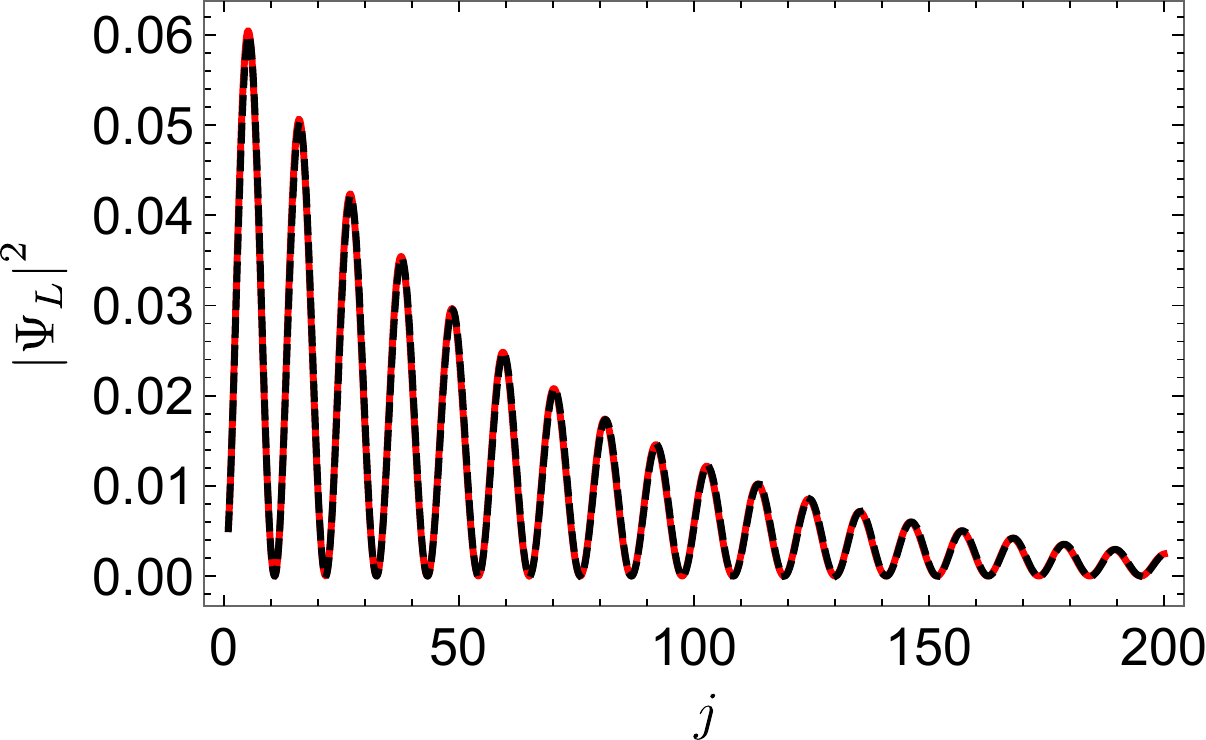}
\caption{The MF probability density $\left|\Psi_L\right|^2$  on the site $j$
obtained numerically (red) and analytically (black). Both approaches agree excellently.
The tight-binding parameters are chosen to be $N= 1000$, $\overline{\alpha}/t_x = 0.3$, $\Delta/t_x = 0.005$, $\mu/t_x = -0.002$, $V_Z/t_x = 0.01$. We note that such the MF probability density can be measured in STM experiments \cite{nadj2014o,pawlak2016p,denis}. } 
\label{fig:Wavefunction}
\end{figure}

In a finite-size nanowire of the length $N a$ ($N \ll N'$), the two MFs split away from zero energy. This energy splitting can be found perturbatively in the framework of the tight-binding model. Here, we represent the Hamiltonian $\overline{H}$ of the finite chain as 
$\overline{H} = H_0^L - H_1$,
where $H_0^L$ is the Hamiltonian of the semi-infinite chain and $H_1$ is the small perturbation that comes from eliminating the hopping between sites $n=N + 1$ and $n= N+2$ and is given by 
\begin{align}
H_1= \sum_{\sigma,\sigma'} 
c^\dag_{N+2,\sigma}\Big[i \overline{\alpha} (\sigma^z)_{\sigma\sigma'}-t_x \delta_{\sigma\sigma'} \Big]c_{N+1,\sigma'}
+{\rm H.c.}
\label{eq:hamiltonian_H1}
\end{align}
As a result, the energy splitting is given by
\begin{align}
\delta\epsilon = \left|\left\langle 0 \left| \gamma_R H_1 \gamma_L^\dag \right| 0 \right\rangle\right| \label{eq:overlap0},
\end{align}
where we used the fact that $ \mathcal{H}_0^L \Psi^L = 0$.  In the Bogoliubov-de-Gennes representation, we arrive  at 
\begin{align}
\left\langle 0 \left| \gamma_R H_1 \gamma_L^\dag \right| 0 \right\rangle= -t_x
\sum_{\sigma}\Big[\left(\chi^R_{N+1, 1, \sigma}\right)^*\chi^L_{N+2, 1, \sigma}\Big]
- i\overline{\alpha}\sum_{\sigma}\Big[\sigma\left(\chi^R_{N+1, 1, \sigma}\right)^*\chi^L_{N+2, 1, \sigma}\Big].
\label{eq:overlap1}
\end{align}

This expression can be significantly simplified further by using the properties of the MF wavefunctions: 
$\chi_{n 1 1}^{L/R} = \left(\chi_{n \bar1 1}^{L/R}\right)^*$ and 
$\chi_{n 1 \bar1}^{L/R} = \left(\chi_{n \bar1 \bar1}^{L/R}\right)^*$.
In addition, in our particular setup, 
$\chi_{n 1 \bar1}^L = i \left(\chi_{n 1 1}^L\right)^*$ and 
$\chi_{n 1 \bar1}^R = -i \left(\chi_{n 1 1}^R\right)^*$. Thus,
\begin{align}
&\delta\epsilon =
2 \left|t_x{\rm Im}\left[\chi_{N+1, 1, 1}^R \left(\chi_{N+2, 1, 1}^L\right)^*\right] - \overline{\alpha} {\rm Re}\left[\chi_{N+1, 1, 1}^R \left(\chi_{N+2, 1, 1}^L\right)^*\right]\right|.
\label{eq:approximation0}
\end{align}
To proceed further, we determine $\chi_{N+2, 1, 1}^L$ and $\chi_{N+1, 1, 1}^R$ from the continuum model. 
Using Eqs.~(10) and (11) of the main text, $\chi_{N+2, 1, 1}^L = \sqrt{a} f(x=(N+2)a)/\sqrt{\mathcal{N}}$ and $\chi_{N+1, 1, 1}^R = \sqrt{a} f^*(x=a)/\sqrt{\mathcal{N}}$. The normalization prefactor $\mathcal{N}$  is defined from the condition $\int_{0}^{+\infty}dx\ \left|\Psi_L(x)\right|^2 = 2$ leading us to
\begin{align}
\mathcal{N} = \xi^e + \xi^i - \frac{4\left(1/ \xi^e + 1/\xi^i\right)}{(1/ \xi^e + 1/\xi^i)^2 +\left(2 k_{so} - \mu/\alpha\right)^2}.
\end{align}
To simplify the final expression, we  introduce the new notation  $ g = |g| e^{i\varphi}\equiv \chi_{N+1, 1, 1}^R \left(\chi_{N+2, 1, 1}^L\right)^* =a f^*(x=a) f^*(x=L)/\mathcal{N}$. 
In this case, Eq.(\ref{eq:approximation0}) can be rewritten as
\begin{align}
\delta\epsilon = 
2 \sqrt{t_x^2+\overline{\alpha}^2} |g| \left|\cos(\varphi + \varphi_0)\right|,
\label{eq:approximation02}
\end{align}
where $\cos \varphi_0 = \overline{\alpha}/\sqrt{t_x^2+\overline{\alpha}^2}$.
Next, we determine $|g|$ and $\varphi$ by performing a Taylor expansion,
\begin{align}
&g=
-\dfrac{a e^{i\varphi_L}}{\mathcal{N}} \left(  e^{-i\left(2 k_{so} - \mu/\alpha\right) a -a / \xi^e} -  e^{-a / \xi^i}\right)\left(  e^{-i\left(2 k_{so} - \mu/\alpha\right) L - L / \xi^e} -  e^{-L / \xi^i}\right),
\label{eq:g2}\\
&g \approx \dfrac{a^2 e^{i\varphi_L}}{\mathcal{N}} \left(i\left(2 k_{so} - \mu/\alpha\right) + 1 / \xi^e - 1 / \xi^i \right)\left(e^{-i\left(2 k_{so} - \mu/\alpha\right) L - L / \xi^e} -  e^{-L / \xi^i}\right),
\label{eq:g3}\\
&g \approx \dfrac{a^2 e^{i\left(\varphi_L + \theta + \theta'\right)}}{\mathcal{N}}\sqrt{\left(1 / \xi^e - 1 / \xi^i\right)^2 + \left(2 k_{so} - \mu/\alpha\right)^2}\sqrt{e^{-2L / \xi^e}+e^{-2L / \xi^i}-2e^{-L / \xi^e}e^{-L / \xi^i}\cos\left[\left(2 k_{so} - \mu/\alpha\right) L\right]}.
\end{align}
The phase of $g$ is given by $\varphi = \varphi_L + \theta + \theta'$,
where
\begin{align}
&\theta = \arctan\left(\dfrac{2 k_{so} - \mu/\alpha}{1 / \xi^e - 1 / \xi^i}\right), \ \
\theta' = \arctan\left(\dfrac{e^{-L/\xi^e}\sin\left[\left(2 k_{so} - \mu/\alpha\right)L\right]}{e^{-L/\xi^i} - e^{-L/\xi^e}\cos\left[\left(2 k_{so} - \mu/\alpha\right)L\right]}\right).
\end{align}
In the strong SOI regime ($\mathcal{N}  \approx \xi^e + \xi^i $, $\varphi_0\approx \pi/2$,  $\theta \approx \pi/2$), we arrive at the following expression for the energy splitting between the two MFs,
\begin{align}
&\delta\epsilon \approx 2\frac{\hbar \upsilon_F}{\xi^e + \xi^i}
\sqrt{e^{-2L / \xi^e}+e^{-2L / \xi^i}-2e^{-L / \xi^e}e^{-L / \xi^i}\cos\left[\left(2 k_{so} - \mu/\alpha\right) L\right]}
 \left|\cos\left(\varphi_L + \theta'\right)\right|\, ,
   \label{eq:approximation09}
\end{align}
where $\xi^i$ depends non-monotonically on the applied magnetic field. If orbital effects of the magnetic field are taken into account, $\xi^i$ first shrinks as a function of magnetic field. However, close to the second topological phase transition, it starts to grow.
Next, we analyze Eq.~(\ref{eq:approximation09}) in two regimes: close and far away from the topological phase transition points.

Away from the topological phase transition points, the exterior gap is the smallest in the system, so $\xi^e \gg \xi^i$. As a result, we arrive at the simplified expression
\begin{align}
\delta\epsilon \approx 2\Delta
e^{-L / \xi^e} \left|\cos\left[\varphi_L - \left(2 k_{so} - \mu/\alpha\right)L\right]\right|.
\end{align}
The amplitude of oscillations, $2\Delta e^{-L / \xi^e}$, stays constant as a function of magnetic field. For $\mu = \mu_0 - \beta V_Z^2$ the period of oscillations in Zeeman energy is given by $\delta V_Z = \pi\alpha / 2 \beta L V_Z$ and stays almost constant if $\delta V_Z \ll V_Z$. However, there is a tendency for shrinking of the period. It should be contrasted with the regime of weak SOI~\cite{rainis2013t}, where the period of oscillations grows  as $\delta V_Z = \pi \hbar \sqrt{2V_Z / m^*} / L$. We note that oscillations in the strong SOI regime arise only due to orbital effects. If we would neglect the shift of the chemical potential caused by the magnetic field via orbital effects, $\mu$ would stay constant as well as the energy splitting $\delta \epsilon$ as a function of the magnetic field.

Close to the phase transition points $\xi^i \gg \xi^e$ and the energy splitting is given by $\delta\epsilon \approx 2\left(\sqrt{V_Z^2 - \mu^2} - \Delta\right)
e^{-L / \xi^i}\left|\cos\left(\varphi_L + \theta'\right)\right|$. On one hand, the amplitude of oscillations is enhanced by the exponential prefactor $e^{-L/\xi^i}$. On the other hand, the prefactor $1/\xi^i$ overtakes the behavior, resulting in the suppression of the splitting as $\xi^i$ diverges. Generally, the region of values of the magnetic field, in which $\xi^i \gg \xi^e$, is very narrow and it is difficult to determine the period of oscillations analytically. However, we observe numerically that 
the splitting between MFs both decays and oscillates as a function of magnetic field close to the second topological phase transition point, see Fig.~\ref{fig:Energy21}.

\begin{figure}[t!] 
\includegraphics[width=0.45\linewidth]{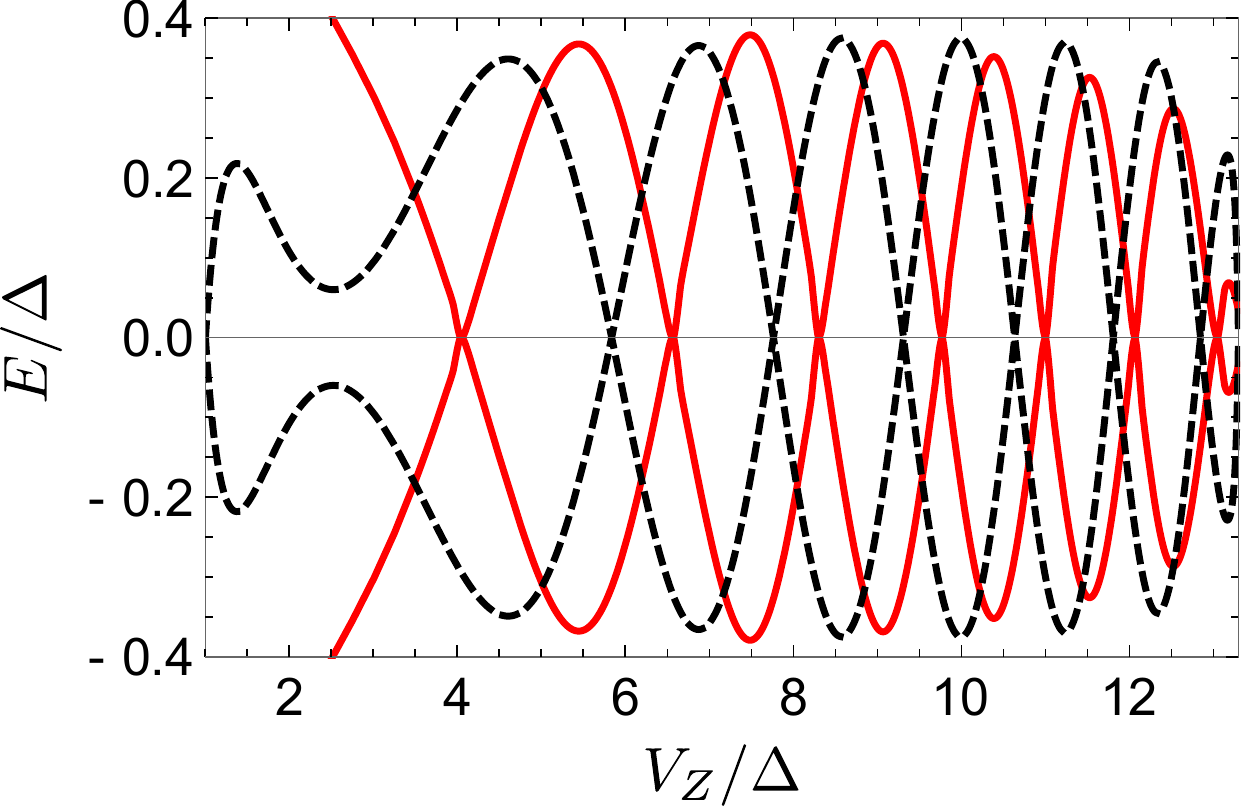}
\caption{The energy splitting between two MFs as a function of applied magnetic field in units of  $V_Z/\Delta$ in the finite-size nanowire $L/a = 302$.
 The results were obtained numerically (red solid line) by exact diagonalization of Eq.~(12) of the main text and analytically (black dashed line) using Eq.~(\ref{eq:approximation09}), see Fig. 5 of the main text. Generally, there is a reasonable agreement between two approaches. The parameters are fixed as $E_{so}/\Delta = 50$ and  $\mu = - 0.075V_Z^2/\Delta$. }
\label{fig:Energy21}
\end{figure}

\section{Linear dependence of the chemical potential on magnetic field: $\mu = \mu_0 - \beta_2 V_Z$}

\begin{figure}[b!] 
\includegraphics[width=0.45\linewidth]{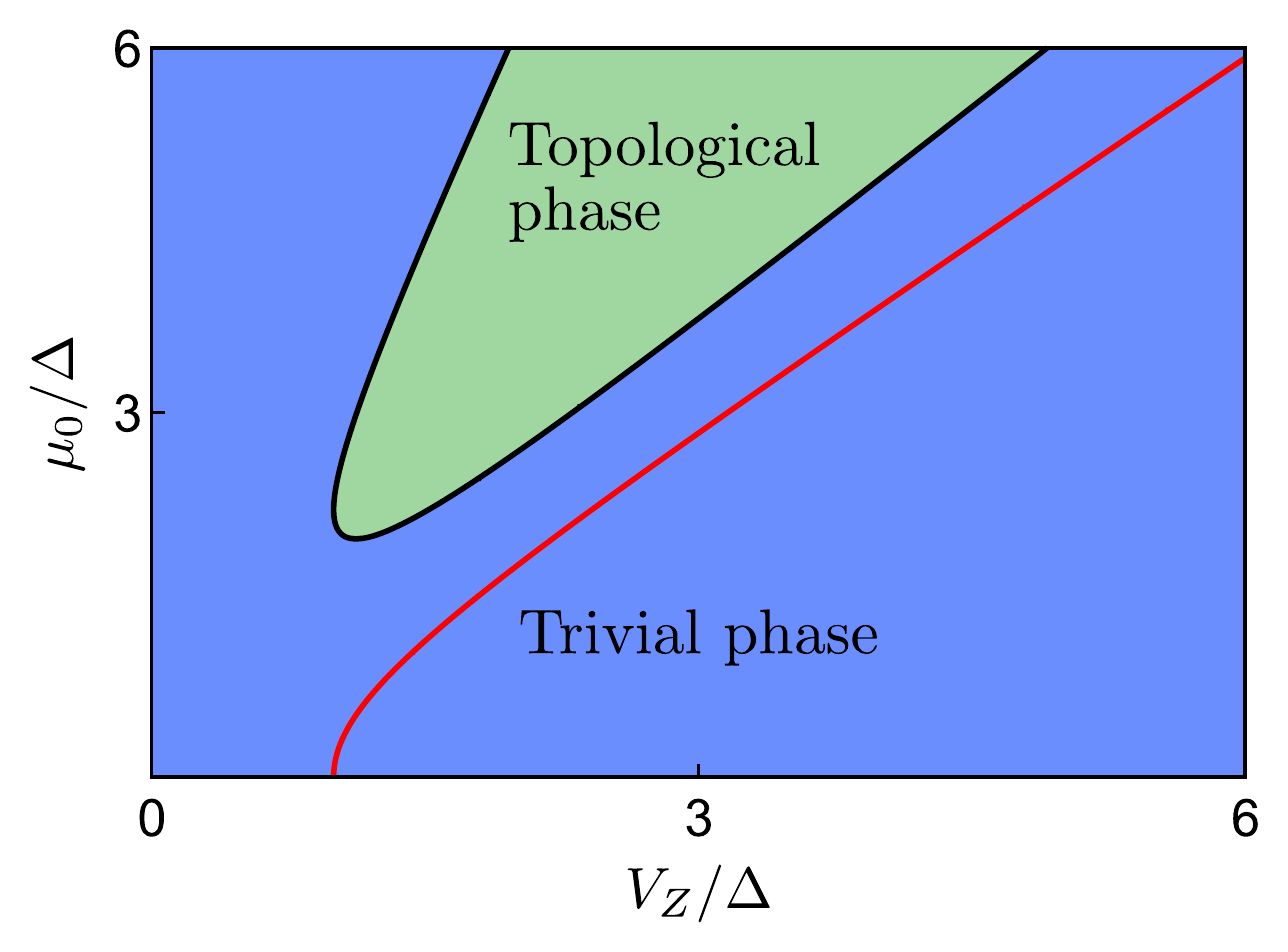}
\caption{Topological phase diagram as a function of applied magnetic field in units of $V_Z/\Delta$ and initial chemical potential $\mu_0/\Delta$ in the case of linear shift of chemical potential ($\beta_2 = 2.2$). The topological (green area) and trivial (blue area) phases are separated by the phase boundary (black line) corresponding to the closing of the bulk gap. The red line indicates the phase boundary in the absence of orbital magnetic effects. Again, the topological phase is shifted towards higher values of initial chemical potential.} 
\label{fig:PhaseDiagram2}
\end{figure}

In this section, we briefly comment on the case if the chemical potential moves linearly as a function of magnetic field. This is the case for degenerate bands and also holds for high values of magnetic fields for all subbands. Thus, we assume that the chemical potential linearly depends on the magnetic field,
\begin{align}
\mu = \mu_0 - \beta_2 V_Z,
\end{align}
where $\mu_0$ is the initial value of the chemical potential and $\beta_2$ is the dimensionless parameter, which is chosen to be positive so that $\beta_2 V_Z>0$. For $1 < \beta_2^2 < 1 + \left(\mu_0/\Delta\right)^2$ the system is in the topological phase if $\mu_0 > 0$ and $V_{Z,-} < V_Z < V_{Z,+}$, where the two critical values of magnetic field are defined as
\begin{align}
V_{Z,\pm} = \dfrac{\beta_2 \mu_0 \pm \sqrt{\mu_0^2 + (1-\beta_2^2)\Delta^2}}{\beta_2^2 - 1}.
\end{align}
In this case, we observe similar oscillations of the energy splitting between two MFs (see Fig.~\ref{fig:Energy_beta2}). Again, the amplitude stays constant for a large range of magnetic fields, $\xi^e \gg \xi^i$. Close to the second topological phase transition point $V_{Z,+}$, the amplitude of oscillations shrinks. Generally, this region of shrinking amplitude gets larger due to the slower dependence of $\mu$ on $V_Z$. The period of oscillations is constant and is given by $\delta V_Z = \pi \alpha / \beta_2 L$.

If $\beta_2^2 < 1$ the system is in the topological phase for $V_Z > V_{Z,-}$ and there is no second topological phase transition.  In the special case when $\beta_2 = 1$, the system is in the topological phase for $V_Z > \left(\Delta^2 + \mu_0^2\right)/2\mu_0$ and $\mu_0>0$.


\begin{figure}[t!] 
\includegraphics[width=0.45\linewidth]{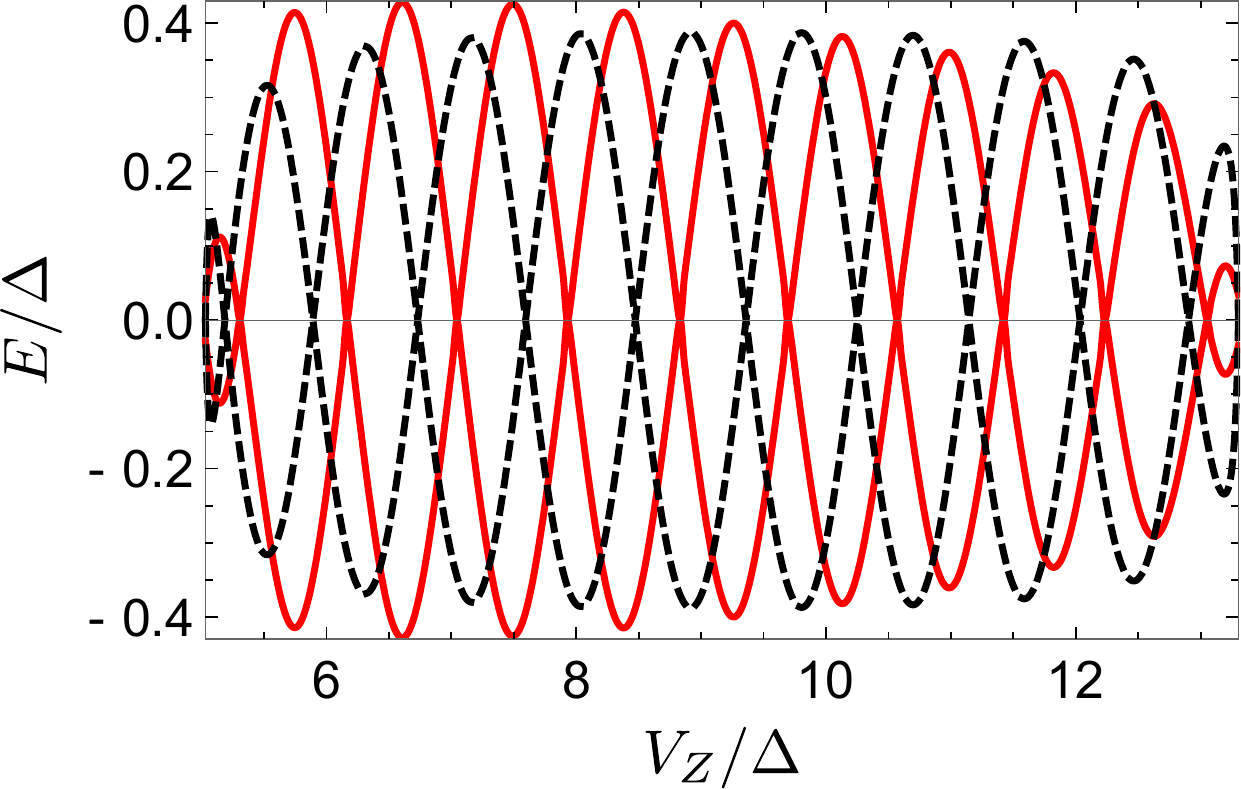}
\caption{The MF energy splitting as a function of applied magnetic field in units of  $V_Z/\Delta$ in a finite-size nanowire $k_{so} L = 151$. The results were obtained numerically (red solid line) by exact diagonalization of Eq.~(12) of the main text and analytically (black dashed line) using Eq.~(\ref{eq:approximation09}). The parameters are chosen as  $E_{so}/\Delta = 50$ and the chemical potential $\mu/\Delta = 16- 2.2 V_Z/\Delta$ is assumed to be linearly shifted due to orbital magnetic effects.  
} 
\label{fig:Energy_beta2}
\end{figure}

\section{Dependence of proximity-induced superconducting gap on magnetic field}

Next, we also include effects of the external magnetic field on the bulk $s$-wave superconductor in the regime of strong coupling between nanowire and bulk superconductor~\cite{reeg2017f}. The proximity-induced gap $\Delta = \Delta_0\sqrt{1 - \left(V_Z / V_Z^c\right)^2}$ is suppressed at the critical field $V_Z^c$, where $\Delta_0$ is the value of the superconducting gap in the absence of magnetic fields. We note that in the weak coupling regime, the proximity induced gap is determined by the tunneling rate between the nanowire and the bulk superconductor and, thus, the proximity-induced gap in the nanowire can be considered independent of the external magnetic field acting on the bulk superconductor
 \cite{sau2010r,potter2011e,kopnin2011p,zyuzin2013c,
takane2013s,van2016c,reeg2017d}.

\subsection{Constant chemical potential $\mu = \mu_0$}

If the bottom of the band is not shifted by orbital effects due to the magnetic field, the chemical potential stays constant $\mu = \mu_0$. The topological criterion is only slightly modified to
\begin{align}
\sqrt{\dfrac{\mu_0^2 +\Delta_0^2}{1 + \left(\Delta_0 / V_Z^c\right)^2}}<V_Z<V_Z^c\, .
\end{align}
In this case, there are no oscillations in the energy splitting [see Fig.~\ref{fig:Energy3}(a)] since the oscillations could only appear due to the shift of the chemical potential with magnetic field. As a result, the amplitude of the energy splitting, first, rapidly increases  and subsequently stays almost constant up to the point where the proximity-induced  gap closes at $V_Z = V_Z^c$.

\begin{figure}[t!] 
\includegraphics[width=0.95\linewidth]{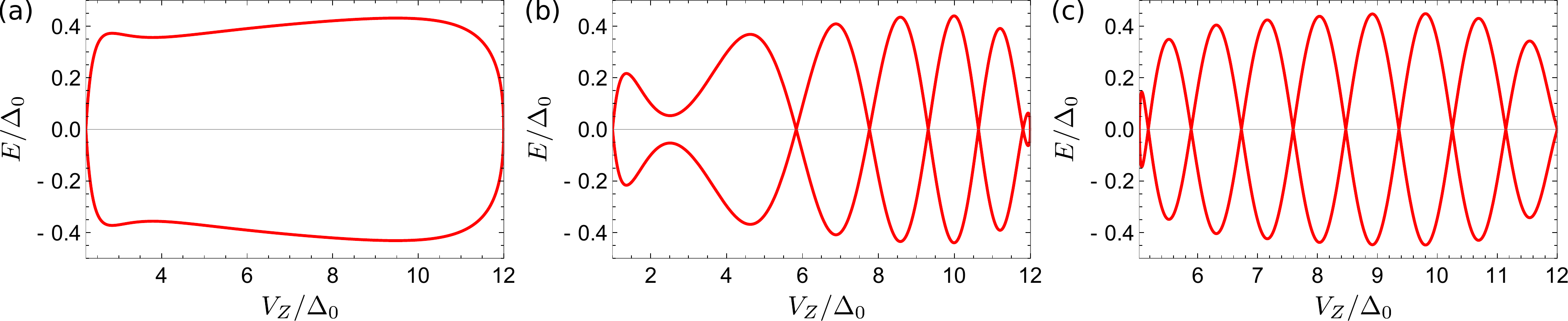}
\caption{The energy splitting between two MFs as a function of applied magnetic field in units of  $V_Z/\Delta_0$ in a finite-size nanowire $k_{so} L = 151$ with $E_{so}/\Delta_0 = 50$. The parameters of the superconducting gap are chosen as follows $\Delta= \Delta_0\sqrt{1 - \left(V_Z / V_Z^c\right)^2}$ with $V_Z^c = 12\Delta_0$, where $V_Z^c$ corresponds to the critical magnetic field $B^c = 2$~T for $\Delta_0 = 0.25$~meV. (a) The chemical potential is fixed to $\mu_0/\Delta_0 = 2$. As expected, if the chemical potential is kept constant, there are no oscillations in the energy splitting. (b) The chemical potential is shifted quadratically due to orbital effects as $\mu/\Delta_0 = -0.075 V_z^2/\Delta_0^2$. 
 (c) The chemical potential is shifted linearly due to orbital effects as $\mu/\Delta_0 = 16 - 2.2 V_z/\Delta_0$.  We note that in both panels (b) and (c), there are oscillations in the energy splitting between MFs and there is a range of magnetic fields for which the MF energy splitting amplitude stays almost constant.} 
\label{fig:Energy3}
\end{figure}


\subsection{Quadratic dependence of the chemical potential $\mu = \mu_0 - \beta V_Z^2$}
Now we consider the chemical potential that is shifted as a quadratic function of magnetic field via orbital effects, $\mu = \mu_0 - \beta V_Z^2$.
If $\mu_0 > -\left(1 + \left(\Delta_0/V_Z^c\right)^2\right)/2\beta$ and $\mu_0 > \left[4\beta^2\Delta_0^2 -\left( 1 + \left(\Delta_0/V_Z^c\right)^2\right)^2\right]/\left[4\beta\left(1 + \left(\Delta_0/V_Z^c\right)^2\right)\right]$ the system is in the topological phase for $V_{Z,-} < V_Z < \text{min}\lbrace V_{Z,+}, V_Z^c\rbrace$, where
\begin{align}
V_{Z,\pm} = \sqrt{\dfrac{\left(1+2\beta\mu_0 + (\Delta_0/V_{Z}^c)^2\right) \pm \sqrt{\left(1+2\beta\mu_0 + (\Delta_0/V_{Z}^c)^2\right)^2 - 4\beta^2(\Delta_0^2 + \mu_0^2)}}{2\beta^2}}.
\end{align}

Again, in the strong coupling regime, the proximity-induced gap $\Delta = \Delta_0\sqrt{1 - \left(V_Z / V_Z^c\right)^2}$ decreases  as the magnetic field is increased. Thus, the localization  length $\xi^e$, which now depends on $V_Z$,  increases with increasing the magnetic field. Away from the topological phase transition point $V_{Z,-}$, the localization length $\xi^e \gg \xi^i$ and the energy splitting first increases with increasing $V_Z$ and then starts to decrease (there is an interplay between the growing prefactor $e^{-L/\xi^e}$ and the decreasing one $\Delta$), see Fig.~\ref{fig:Energy3}(b). Close to the topological phase transition point $V_{Z,-}$, the localization  length $\xi^i \gg \xi^e$, so the energy splitting is increasing in this very narrow region.

\subsection{Linear dependence of the chemical potential  $\mu = \mu_0 - \beta_2 V_Z$}
Finally, we consider the chemical potential that linearly depends on the magnetic field, $\mu = \mu_0 - \beta_2 V_Z$.
For $1 + (\Delta_0/V_{Z}^c)^2 < \beta_2^2 < \left(1 + (\mu_0/\Delta_0)^2\right)\left(1 + (\Delta_0/V_{Z}^c)^2 \right)$ the system is in the topological phase if $\mu_0>0$ and $V_{Z,-} < V_Z < \text{min}\lbrace V_{Z,+}, V_Z^c\rbrace$, where
\begin{align}
V_{Z,\pm} = \dfrac{\beta_2 \mu_0 \pm \sqrt{\beta_2^2\mu_0^2 + (\Delta_0^2 + \mu_0^2)\left(-\beta_2^2 + 1 + (\Delta_0/V_{Z}^c)^2 \right)}}{\left(\beta_2^2 - 1 - (\Delta_0/V_{Z}^c)^2 \right)}.
\end{align}
 In this case, we observe oscillations of the energy splitting that have similar features as the ones obtained for the quadratic dependence of the chemical potential [see Fig.~\ref{fig:Energy3}(c)].
For $\beta_2^2 < 1 + (\Delta_0/V_{Z}^c)^2$ the system is in the topological phase if $V_{Z, -}< V_Z < V_Z^c$. In the special case $\beta_2^2 = 1 + (\Delta_0/V_{Z}^c)^2$, the system is in the topological phase if $\mu_0>0$ and  $\left(\Delta_0^2 + \mu_0^2\right)/2\mu_0\sqrt{1 + (\Delta_0/V_{Z}^c)^2} < V_Z < V_Z^c$.

\end{widetext}
\end{document}